\documentstyle[preprint,aps]{revtex}

\input psfig.sty

\begin{document}
\draft

\title{Relativistic direct Urca processes in cooling neutron stars}
\author{L. B. Leinson$^{a,b}$ and A. P\`{e}rez$^{a}$}
\address{$^{a}$Departamento de F\`{\i}sica Te\`{o}rica, Universidad de
Valencia, 46100 Burjassot (Valencia), Spain \\
$^{b}$Institute of Terrestrial Magnetism, Ionosphere and Radio Wave
Propagation RAS, 142090 Troitsk, Moscow Region, Russia}

\maketitle

\begin{abstract}
We derive a relativistic expression for neutrino energy losses caused by the
direct Urca processes in degenerate baryon matter of neutron stars. We use
two different ways to calculate the emissivity caused by the reactions to
our interest. First we perform a standard calculation by Fermi's ''golden''
rule. The second calculation, resulting in the same expression, is performed
with the aid of polarization functions of the medium. Our result for
neutrino energy losses strongly differs from previous non-relativistic
results. We also discuss nonconservation of the baryon vector current in
reactions through weak charged currents in the medium, when the asymmetry
between protons and neutrons is considered. The above effects, not discussed
in the literature before, substantially modify the polarization functions
responsible for the induced weak charged currents in baryon matter. 
\end{abstract}
\widetext

In numerical simulations, neutrino cooling of the massive core of a neutron
star, with the standard nuclear composition, is governed by the direct Urca
processes $n\rightarrow p+e^{-}+\bar{\nu}_{e}$, $\ \ \ p+e^{-}\rightarrow
n+\nu _{e}$. The important role of these reactions in the rapid cooling of
neutron stars has been first pointed out by Boguta \cite{B81}. Later
Lattimer et al. \cite{Lat91} have suggested a simple formula for neutrino
energy losses caused by the above processes in degenerate nuclear matter
under $\beta $-equilibrium, which exhibits a threshold of proton
concentration necessary for the direct Urca processes to operate\footnote{%
Prakash et al. \cite{Prak92} have found that an admixture of $\Lambda $
hyperons opens the direct Urca processes in baryon matter for an arbitrary
proton concentration.}. The above estimates have been made in a
non-relativistic manner, assuming that the participating particles are free.
Some improvement, which partially accounts for strong interactions, was made
by replacing the baryon masses with their effective values.

Actually, the superthreshold proton fraction in the core of neutron stars
appears at large densities, when Fermi momenta of participating nucleons are
of the order, or larger, than their effective masses. The total
four-momentum of the final lepton and antineutrino is time-like; therefore,
in the free relativistic gas, the energy-momentum conservation requires a
large difference in the effective masses of protons and neutrons $%
M_{n}^{\ast }-M_{p}^{\ast }\sim 10^{2}MeV$ that is unlikely to appear. Thus,
in the relativistic regime, the energy conservation can be fulfilled only by
taking into account the difference in the potential energies of proton and
neutron. A further simplification was made by neglecting the proton recoil.
This is not justified in the relativistic regime, because the momentum of
the final lepton is of the order of the proton effective mass. Both the
proton recoil and the difference in the proton-neutron potential energies
strongly modify the emissivity of the direct Urca reactions {\em even in the
Mean Field Approximation}

In the present paper we derive a relativistic expression for neutrino energy
losses caused by the direct Urca processes by taking into account the above
effects. We use two different ways to calculate the emissivity caused by the
reactions to our interest. First we perform a standard calculation by
Fermi's ''golden'' rule. The second calculation, resulting in the same
expression, is performed with the aid of polarization functions of the
medium. In addition to allow us to check the result obtained by the previous
method, this second method shows some important details, which must be taken
into account in the next step (when RPA corrections are included).

Relativistic polarization functions of hot baryon matter have been
investigated earlier by several authors (see e.g. \cite{SMS89}, \cite{HPH95}%
, \cite{RPLP99} and references therein) having made analytic computations
for the case when the particle-hole excitations correspond to identical
baryons. The study of interactions through charged weak-currents requieres
considering polarization functions of another type, in which the virtual
particle and the hole correspond to baryons of different kinds. As we show
below, the difference in potential energies of participating particles leads
to non-conservation of the vector transition current of baryons in the
direct Urca reactions. This effect dramatically modifies the corresponding
polarization functions of the medium and must be taken into account in order
to obtain the correct result.

According to mean-field theory, the medium-modified energy of a relativistic
baryon of kind $B$ is similar to the energy of a single particle with
effective Dirac mass $M_{B}^{\ast }$ in the effective potential $U_{B}$ of
self-consistently generated meson fields\footnote{%
In what follows we use the system of units $\hbar =c=1$ and the Boltzmann
constant $k_{B}=1$.}. The effective mass, as well as the effective
potential, arises due to interactions of the fields. We denote by $%
\varepsilon _{B}\left( {\bf p}\right) =\sqrt{{\bf p}^{2}+M_{B}^{\ast 2}}$
the baryon kinetic energy. The single-particle energy of a baryon with
momentum ${\bf p}$ is $E_{B}\left( {\bf p}\right) =\varepsilon _{B}\left( 
{\bf p}\right) +U_{B}$, so that the individual Fermi distributions are of
the form 
\begin{equation}
f_{B}\left( \varepsilon _{B}\right) =\frac{1}{\exp \left( \left( \varepsilon
_{B}+U_{B}-\mu _{B}\right) /T\right) +1},  \label{dist}
\end{equation}%
where $\mu _{B}$ is the chemical potential of $B$-kind baryons.

The low-energy Lagrangian of baryon interaction with the lepton field can be
written in a point-like approach (summation over repeated Greek indexes is
assumed) 
\begin{equation}
{\cal L}_{{\rm weak}}=\frac{G_{F}C}{\sqrt{2}}j^{\alpha }J_{\alpha },
\label{H}
\end{equation}%
where $G_{F}$ is the Fermi\ weak coupling constant, and the Cabibbo factor $%
C=\cos \theta _{C}=0\allowbreak .\,\allowbreak 973$ for change of
strangeness $\Delta S=0$ and $C=\sin \theta _{C}$ for $\Delta S=1$. For the
direct Urca process involving baryons $B_{1}$ and $B_{2}$ 
\begin{equation}
B_{1}\rightarrow B_{2}+l+\bar{\nu}_{l},  \label{decay}
\end{equation}%
the lepton and baryon weak charged currents are, respectively: 
\begin{equation}
j^{\alpha }=\bar{u}_{l}\gamma ^{\alpha }\left( 1-\gamma _{5}\right) \nu
_{l},\ \ \ \ \ \ J_{\alpha }=\bar{\psi}_{2}\gamma _{\alpha
}(C_{V}-C_{A}\gamma _{5})\psi _{1}.  \label{wc}
\end{equation}%
Here $\psi _{1}$ and $\psi _{2}$ stand for the initial and final baryon
fields; $C_{V}$ and $C_{A}$ \ are the corresponding vector and axial-vector
coupling constants, respectively. In what follows we consider massless
neutrinos of energy and momentum $q_{1}=\left( \omega _{1},{\bf k}%
_{1}\right) $ with $\omega _{1}=k_{1}.$ The energy and momentum of the final
lepton of mass $m_{l}$ is denoted as $q_{2}=\left( \omega _{2},{\bf k}%
_{2}\right) $ with $\omega _{2}=\sqrt{k_{2}^{2}+m_{l}^{2}}$.

We consider the total energy which is emitted into neutrino and antineutrino
per unit volume and time. Within $\beta $-equilibrium, the inverse reaction $%
B_{2}+l\rightarrow B_{1}+\nu _{l}$ corresponding to the capture of a lepton $%
l$, gives the same emissivity as the decay (\ref{decay}), but in neutrinos.
Thus, the total energy loss $Q$ for the Urca processes is twice more than
that caused by $\beta $-decay (\ref{decay}). Taking this into account by
Fermi's ''golden'' rule we have%
\begin{eqnarray}
Q &=&2\frac{G_{F}^{2}C^{2}}{2}\int \frac{%
d^{3}k_{2}d^{3}k_{1}d^{3}p_{2}d^{3}p_{1}}{2\omega _{2}2\omega
_{1}2\varepsilon _{2}2\varepsilon _{1}(2\pi )^{12}}\left| {\cal M}%
_{fi}\right| ^{2}\omega _{1}\,f_{1}\left( 1-f_{2}\right) \left(
1-f_{l}\right)  \nonumber \\
&&\times \left( 2\pi \right) ^{4}\delta \left( \varepsilon
_{1}+U_{1}-\varepsilon _{2}-U_{2}-\omega _{1}-\omega _{2}\right) \delta
\left( {\bf p}_{1}-{\bf p}_{2}-{\bf k}_{1}-{\bf k}_{2}\right) ,  \label{gold}
\end{eqnarray}%
where the square of the matrix element of the reaction (\ref{decay}) summed
over spins of initial and final particles has the following form 
\begin{eqnarray}
\left| {\cal M}_{fi}\right| ^{2} &=&64\left[ \left(
C_{A}^{2}-C_{V}^{2}\right) M_{1}^{\ast }M_{2}^{\ast }\left(
q_{1}q_{2}\right) +\left( C_{A}-C_{V}\right) ^{2}\left( q_{1}P_{2}\right)
\left( q_{2}P_{1}\right) \right.  \nonumber \\
&&\left. +\left( C_{A}+C_{V}\right) ^{2}\left( q_{1}P_{1}\right) \left(
q_{2}P_{2}\right) \right]  \label{MatrEl}
\end{eqnarray}%
with $P_{1}=\left( \varepsilon _{1,}{\bf p}_{1}\right) $ and $P_{2}=\left(
\varepsilon _{2,}{\bf p}_{2}\right) $. Antineutrinos are assumed to be
freely escaping. The distribution function of initial baryons $B_{1}$ as
well as blocking of final states of the baryon $B_{2}$ and the lepton $l$
are taken into account by the Pauli blocking-factor $\,f_{1}\left(
1-f_{2}\right) \left( 1-f_{l}\right) $ with 
\begin{equation}
f_{l}\left( \omega _{2}\right) =\frac{1}{\exp \left( \omega _{2}-\mu
_{l}\right) /T+1}  \label{fl}
\end{equation}%
being the Fermi-Dirac distribution function of leptons with the chemical
potential $\mu _{l}$. By neglecting the chemical potential of escaping
neutrinos, we can write the condition of chemical equilibrium as $\mu
_{l}=\mu _{1}-\mu _{2}$. We assume degenerate matter. Then by the use of the
energy conservation equation $\varepsilon _{1}+U_{1}=\varepsilon
_{2}+U_{2}+\omega _{2}+\omega _{1}$, and taking the total energy of the
final lepton and antineutrino as $\omega _{2}+\omega _{1}=\mu _{l}+\omega
^{\prime }$ we obtain the identity 
\begin{eqnarray}
&&\,f_{1}\left( \varepsilon _{1}\right) \left( 1-f_{2}\left( \varepsilon
_{2}\right) \right) \left( 1-f_{l}\left( \omega _{2}\right) \right) 
\nonumber \\
&\equiv &f_{1}\left( \varepsilon _{1}\right) \left( 1-f_{1}\left(
\varepsilon _{1}-\omega ^{\prime }\right) \right) \left( 1-f_{l}\left( \mu
_{l}+\omega ^{\prime }-\omega _{1}\right) \right) ,  \label{block}
\end{eqnarray}%
where $\omega ^{\prime }\sim T$ because the energy exchange in the direct
Urca reaction goes naturally on the temperature scale $\sim T$. Due to the
strong degeneracy of the medium, the main contribution to the integral (\ref%
{gold}) comes from narrow regions of momentum space near the corresponding
Fermi surfaces. Since the neutrino energy is $\omega _{1}\sim T$, and the
neutrino momentum $k_{1}\sim T$ is much smaller than the momenta of other
particles, we can neglect the neutrino contributions in the energy-momentum
conserving delta-functions. Then%
\begin{eqnarray}
&&\delta \left( \varepsilon _{1}+U_{1}-\varepsilon _{2}-U_{2}-\omega
_{1}-\omega _{2}\right) \delta \left( {\bf p}_{1}-{\bf p}_{2}-{\bf k}_{1}-%
{\bf k}_{2}\right)  \nonumber \\
&\simeq &\frac{\varepsilon _{2}}{p_{1}k_{2}}\delta \left( \cos \theta -\frac{%
1}{2p_{1}k_{2}}\left( p_{1}^{2}-p_{2}^{2}+k_{2}^{2}\right) \right) \delta
\left( {\bf p}_{1}-{\bf p}_{2}-{\bf k}_{2}\right) ,  \label{delfun}
\end{eqnarray}%
where the $\theta $ is the angle between the momentum ${\bf p}_{1}$ of the
initial baryon and the momentum ${\bf k}_{2}$ of the final lepton. Now
integrations over ${\bf p}_{2}$ momenta and all solid angles can be easily
done. As mentioned above, we can replace the momenta of particles by the
corresponding Fermi momenta, i.e. $p_{1}=p_{F1}$, $p_{2}=p_{F2}$, $%
k_{2}=p_{Fl}$ in all smooth functions of energy and momentum under the
integral. Then the remaining integration is reduced to the following%
\begin{eqnarray}
&&\int d\omega _{1}\omega _{1}^{3}d\omega ^{\prime }d\varepsilon
_{1}\,f_{1}\left( \varepsilon _{1}\right) \left( 1-f_{1}\left( \varepsilon
_{1}-\omega ^{\prime }\right) \right) \left( 1-f_{l}\left( \mu _{l}+\omega
^{\prime }-\omega _{1}\right) \right)  \nonumber \\
&\simeq &\int_{-\infty }^{\infty }d\omega ^{\prime }\frac{\omega ^{\prime }}{%
\exp \omega ^{\prime }/T-1}\int_{0}^{\infty }d\omega _{1}\frac{\omega
_{1}^{3}}{1+\exp \left( \omega _{1}-\omega ^{\prime }\right) /T}=\frac{457}{%
5040}\pi ^{6}T^{6}.  \label{int}
\end{eqnarray}%
Finally, to the lowest order in $T/\mu _{l}$ we obtain the following
neutrino emissivity\footnote{
We have corrected two misprints appearing in the
published version of our paper (see Eq. (11) in ref. \cite{PLB}).}:%
\begin{eqnarray}
Q &=&\,\frac{457\pi }{10\,080}G_{F}^{2}C^{2}T^{6}\left[ C_{V}C_{A}\left(
\left( \varepsilon _{F1}+\varepsilon _{F2}\right) p_{Fl}^{2}-\left(
\varepsilon _{F1}-\varepsilon _{F2}\right) \left(
p_{F1}^{2}-p_{F2}^{2}\right) \right) \right.  \nonumber \\
&&+2C_{A}^{2}\mu _{l}M_{1}^{\ast }M_{2}^{\ast }+\left(
C_{V}^{2}+C_{A}^{2}\right) \left( \mu _{l}\left( 2\varepsilon
_{F1}\varepsilon _{F2}-M_{1}^{\ast }M_{2}^{\ast }\right) +\varepsilon
_{F1}p_{Fl}^{2}\right.  \nonumber \\
&&\left. \left. -\frac{1}{2}\left( \varepsilon _{F1}+\varepsilon
_{F2}\right) \left( p_{F1}^{2}-p_{F2}^{2}+p_{Fl}^{2}\right) \right) \right]
\Theta \left( p_{Fl}+p_{F2}-p_{F1}\right)  \label{QMFA}
\end{eqnarray}%
with $\Theta \left( x\right) =1$ if $x\geq 0$ and zero otherwise. When the
baryon and lepton momenta are at their individual Fermi surfaces, the $%
\delta $- function (\ref{delfun}) contributes only if $p_{F2}+p_{Fl}>p_{F1}$%
. This ''triangle'' condition required by the step-function in Eq. (\ref%
{QMFA}) defines the above mantioned threshold for the direct Urca reactions.

Eq.(\ref{QMFA}) can be obtained also by the use of the medium response
functions. By the fluctuation-dissipation theorem, the total emissivity can
be written as: 
\begin{equation}
Q=2\frac{G_{F}^{2}C^{2}}{2}\int \;\frac{\omega _{1}\ \left( 1-f_{l}\left(
\omega _{2}\right) \right) \,2\mathop{\rm Im}\left[ W_{R}^{\alpha \beta }%
{\rm L}_{\alpha \beta }\right] }{\exp \left( q_{0}-\mu _{1}+\mu _{2}\right)
/T-1}\frac{d^{3}k_{2}}{(2\pi )^{3}}\frac{d^{3}k_{1}}{(2\pi )^{3}},
\label{Qnu}
\end{equation}%
where the integration goes over the phase volume of the final lepton and
antineutrino of total energy $q_{0}=\omega _{1}+\omega _{2}$ and total
momentum ${\bf q=k}_{1}+{\bf k}_{2}$. The final-state blocking of the
outgoing lepton is taken into account by the Pauli blocking-factor $\left(
1-f_{l}\left( \omega _{2}\right) \right) $. The factor $\left[ \exp \left(
q_{0}-\mu _{1}+\mu _{2}\right) /T-1\right] ^{-1}$ in Eq. (\ref{Qnu}) arises
due to the fact that the baryons $B_{1}$ and $B_{2}$ are in thermal
equilibrium at temperature $T$ and in chemical equilibrium with chemical
potentials $\mu _{1}$ and $\mu _{2}$, respectively. ${\rm L}^{\alpha \beta } 
$ is defined here by 
\begin{equation}
{\rm L}^{\alpha \beta }=\frac{8}{2\omega _{2}2\omega _{1}}\left(
k_{1}^{\alpha }k_{2}^{\beta }+k_{2}^{\alpha }k_{1}^{\beta }-\left(
k_{1}\cdot k_{2}\right) g^{\alpha \beta }-i\epsilon ^{\alpha \beta \gamma
\delta }\left( k_{1}\right) _{\gamma }\left( k_{2}\right) _{\delta }\right) ,
\label{Ltens}
\end{equation}%
and $W_{R}^{\alpha \beta }$ is the retarded weak-polarization tensor of the
medium. The imaginary part of the retarded polarization is related to that
of the causal (or time ordered) polarization $W_{\alpha \beta }$ as follows 
\begin{equation}
\mathop{\rm Im}W_{R}^{\alpha \beta }=\tanh \left( \frac{q_{0}-\mu _{1}+\mu
_{2}}{2T}\right) \mathop{\rm Im}W^{\alpha \beta }.  \label{Ret}
\end{equation}%
Since the baryon weak-current includes vector and axial-vector
contributions, to the lowest order in weak interactions, the polarization
tensor consists on those diagrams which have ends at the weak vertex $\left(
C_{V}\gamma ^{\alpha }-C_{A}\gamma ^{\alpha }\gamma _{5}\right) $. Thus, we
can take the weak polarization tensor as the sum of vector-vector,
axial-axial, and mixed axial-vector pieces: 
\begin{equation}
W^{\alpha \beta }=C_{V}^{2}\Pi ^{\alpha \beta }+C_{A}^{2}\Pi _{A}^{\alpha
\beta }-2C_{V}C_{A}\Pi _{VA}^{\alpha \beta }.  \label{Tensor}
\end{equation}

The vector-vector polarization tensor is of the following general form 
\begin{eqnarray}
\Pi ^{\alpha \beta } &=&-i\int \frac{d^{3}pdP^{0}}{(2\pi )^{4}}\,\mathop{\rm
Tr}\left[ \,\,G_{1}\left( P^{0}-U_{1}+q_{0},{\bf p+q}\right) \,\gamma
^{\alpha }G_{2}\left( P^{0}-U_{2},{\bf p}\right) \gamma ^{\beta }\right] , 
\nonumber \\
&&  \label{PiV}
\end{eqnarray}%
were $Q^{\mu }=\left( q_{0},{\bf q}\right) $ is the total four-momentum
transfer, and the baryon propagators $G_{1}$ and $G_{2}$ correspond to
different baryons $B_{1}$ and $B_{2}$. In the following we consider a not
too large momentum transfer and not too high temperatures, permitting us to
neglect antibaryons in the system. Then the in-medium baryon propagator can
be written as: 
\begin{equation}
{\hat{G}}_{B}(P)=\left( \gamma P_{B}^{\ast }+M_{B}^{\ast }\right) \left[ 
\frac{1}{P_{B}^{\ast 2}-M_{B}^{\ast 2}+i0}+\frac{\pi i}{\varepsilon _{B}}%
\delta \left( P_{B}^{\ast 0}-\varepsilon _{B}\right) f_{B}\left( \varepsilon
_{B}\right) \right] .  \label{G}
\end{equation}%
The first term in Eq. (\ref{G}) describes propagation of free baryons, and
the second term includes the Pauli principle restrictions\footnote{%
Notice that this form of the baryon propagator takes into account not only
the Pauli principle, but also partially strong interactions, which are
included by means of the effective mass of the baryon and the
self-consistent nuclear field.}. Thus, the polarization tensors is the sum
of a Feynman piece and an explicitly density-dependent piece. The
energy-momentum transferred in the considered processes are not too large ($%
q_{0}^{2}-{\bf q}^{2}\simeq m_{l}^{2}$, with $m_{l}$ being the lepton mass),
therefore we neglect the vacuum contribution and consider only the
density-dependent part of polarization functions that are related to virtual
excitations of a particle-hole type, with $P_{B}^{\ast }=\left( P^{0}-U_{B},%
{\bf p}\right) $.

By a simple change of the integration variable $p_{0}=P_{0}-U_{2}$, Eq. (\ref%
{PiV}) can be reduced to the standard form 
\begin{equation}
\Pi ^{\alpha \beta }=-i\int \frac{d^{4}p}{(2\pi )^{4}}\mathop{\rm Tr}\left[
\,G_{1}(p+\tilde{Q})\,\gamma ^{\alpha }\,G_{2}(p)\gamma ^{\beta }\right] ,
\label{Piv}
\end{equation}%
with $p^{\mu }=\left( p_{0},{\bf p}\right) $ but with the effective momentum
transfer given by 
\begin{equation}
\tilde{Q}^{\mu }=\left( \tilde{q}_{0},{\bf q}\right) ,\,\ \ \ \ \ \tilde{q}%
_{0}=q_{0}-U_{1}+U_{2}.  \label{qeff}
\end{equation}%
Obviously, the polarization functions depend explicitly on $\tilde{q}_{0}$
instead of the total energy transfer $q_{0}$. The only known generalization
(see refs.\cite{HPH95}, \cite{RPLP99}) of the polarization functions to the
case of different baryons in the polarization loop have been made by the
ansatz that the general form of the tensor (\ref{Piv}) should be a sum of
longitudinal and transverse components. Such an approach (which is valid in
vacuum under the assumption of isotopic invariance of nucleons) is invalid
for a system of interacting baryons, because in a medium the isovector
current, caused by conversion of the baryon $B_{1}\rightarrow $ $B_{2}$, is
not conserved.

Let's examine the nonconservation of the baryon current by contracting
together the vector-vector polarization tensor and the effective
four-momentum transfer 
\begin{equation}
\tilde{Q}_{\alpha }\Pi ^{\alpha \beta }=-i\int \frac{d^{4}p}{(2\pi )^{4}}%
\mathop{\rm Tr}\left[ \,G_{1}(p+\tilde{Q})\,\left( \gamma \tilde{Q}\right)
\,G_{2}(p)\gamma ^{\beta }\right] .  \label{cont}
\end{equation}%
The baryon propagator (\ref{G}) obeys the equation $\left( \gamma
p-M_{B}^{\ast }\right) G_{B}(p\dot{)}=1$ with $1$ being the identity matrix.
By taking this into account and using the following identity%
\begin{eqnarray}
\tilde{q}_{0}\gamma _{0}-{\bf q\gamma } &{\bf =}&\left[ \left( p_{0}{\bf +}%
\tilde{q}_{0}\right) \gamma _{0}-\left( {\bf p+q}\right) {\bf \gamma }%
-M_{1}^{\ast }\right] -\left[ p_{0}\gamma _{0}-{\bf p\gamma }-M_{2}^{\ast }%
\right]  \nonumber \\
&&+\left( M_{1}^{\ast }-M_{2}^{\ast }\right)  \label{iden}
\end{eqnarray}%
we obtain from Eq. (\ref{cont})%
\begin{eqnarray}
\tilde{Q}_{\alpha }\Pi ^{\alpha \beta } &=&-i\int \frac{d^{4}p}{(2\pi )^{4}}%
\mathop{\rm Tr}\,G_{2}(p)\gamma ^{\beta }+i\int \frac{d^{4}p}{(2\pi )^{4}}%
\mathop{\rm Tr}G_{1}(p+\tilde{Q})\gamma ^{\beta }\,  \nonumber \\
&&+\left( M_{1}^{\ast }-M_{2}^{\ast }\right) \,\Pi _{M}^{\beta },  \label{c1}
\end{eqnarray}%
where $\Pi _{M}^{\alpha }$ is the mixed vector-scalar polarization, given by%
\begin{equation}
\Pi _{M}^{\beta }\left( \tilde{Q}\right) =-i\int \frac{d^{4}p}{(2\pi )^{4}}%
\mathop{\rm Tr}\left[ \,G_{1}(p+\tilde{Q})\,\,G_{2}(p)\gamma ^{\beta }\,%
\right] .  \label{Pimix}
\end{equation}%
We can make the change $p+\tilde{Q}\rightarrow p$ in the second integral of
Eq.(\ref{c1}). Then, using the explicit form (\ref{G}) of the baryon
propagators, we obtain%
\begin{equation}
\tilde{Q}_{\alpha }\Pi ^{\alpha \beta }=\delta _{\beta 0}\,\left(
n_{2}-n_{1}\,\right) +\left( M_{1}^{\ast }-M_{2}^{\ast }\right) \,\Pi
_{M}^{\beta },  \label{cnc}
\end{equation}%
where $n_{1}$ and $n_{2}$ are the number densities of baryons $B_{1}$ and $%
B_{2}$, respectively, and $\delta _{\beta 0}$ $=1$ if $\beta =0$ and zero
otherwise. Eq. (\ref{cnc}) can be equivalently written as%
\begin{equation}
q^{0}\Pi ^{0\beta }-q^{j}\Pi ^{j\beta }=\left( U_{1}-U_{2}\right) \Pi
^{0\beta }-\delta _{\beta 0}\,\left( n_{1}-n_{2}\,\right) +\left(
M_{1}^{\ast }-M_{2}^{\ast }\right) \,\Pi _{M}^{\beta }.  \label{cnc1}
\end{equation}

Thus, in a medium, the baryon vector current is conserved only in the case
of symmetric nuclear matter $\left( n_{1}=n_{2}\right) $ and isotopic
invariance of the baryons $\left( M_{1}^{\ast }=M_{2}^{\ast }\right) $,
which together with $U_{1}=U_{2}$ provide $q^{0}\Pi ^{0\beta }-q^{j}\Pi
^{j\beta }=0$. If at least one of the above conditions is not fulfilled,
then $Q_{\alpha }\Pi ^{\alpha \beta }\neq 0$. In particular, this means that
the vector-vector polarization tensor for charged currents can not be
written in terms of longitudinal and transverse components, as assumed in
refs. (\cite{HPH95}, \cite{RPLP99}). In a frame of reference where $Q=\left(
q_{0},0,0,q\right) $, the vector-vector polarization tensor (\ref{Piv}) has
four independent components $\Pi ^{00}$, $\Pi ^{03}=\Pi ^{30}$, $\Pi ^{33}$,
and $\Pi ^{11}=\Pi ^{22}$.

The axial-axial tensor 
\begin{equation}
\Pi _{A}^{\alpha \beta }\left( q_{0},{\bf q}\right) =-i\int \frac{d^{4}p}{%
(2\pi )^{4}}\mathop{\rm Tr}\,\left[ G_{1}(p+\tilde{Q})\gamma ^{\alpha
}\,\gamma _{5}\,G_{2}(p)\gamma ^{\beta }\,\gamma _{5}\right]  \label{Pia}
\end{equation}%
as well as the vector-axial%
\begin{equation}
\Pi _{VA}^{\alpha \beta }\left( q_{0},{\bf q}\right) =-i\int \frac{d^{4}p}{%
(2\pi )^{4}}\mathop{\rm Tr}\,\left[ \,G_{1}(p+\tilde{Q})\gamma ^{\alpha
}\,\,G_{2}(p)\gamma ^{\beta }\,\gamma _{5}\right]  \label{Piva}
\end{equation}%
are of the standard form: 
\begin{equation}
\Pi _{A}^{\mu \nu }=\Pi ^{\mu \nu }+\Pi _{A}g^{\mu \nu },  \label{Aten}
\end{equation}%
\begin{equation}
\Pi _{VA}^{\mu \nu }=\Pi _{VA}\frac{Q_{\lambda }}{q}i\epsilon ^{\mu \nu
\lambda 0}.  \label{VAten}
\end{equation}

Our calculation of the integrals yields the following imaginary components
of the retarded polarization tensors: 
\begin{equation}
\mathop{\rm Im}\Pi _{R}^{00}=\frac{1}{2\pi q}\left[ I_{2}+\tilde{q}_{0}I_{1}+%
\frac{1}{4}\left( t-\left( M_{1}^{\ast }-M_{2}^{\ast }\right) ^{2}\right)
I_{0}\right] ,  \label{Pi00}
\end{equation}%
\begin{eqnarray}
\mathop{\rm Im}\Pi _{R}^{03} &=&\mathop{\rm Im}\Pi _{R}^{30}=\frac{1}{2\pi
q^{2}}\left[ \tilde{q}_{0}I_{2}-\frac{1}{2}\left( M_{1}^{\ast 2}-M_{2}^{\ast
2}-2\tilde{q}_{0}^{2}\right) I_{1}\right.  \nonumber \\
&&\left. +\frac{\tilde{q}_{0}}{4}\left( t-M_{1}^{\ast 2}+M_{2}^{\ast
2}\right) I_{0}\right] ,  \label{Pi03}
\end{eqnarray}%
\begin{eqnarray}
\mathop{\rm Im}\Pi _{R}^{33} &=&\frac{1}{2\pi q^{3}}\left[ \tilde{q}%
_{0}^{2}I_{2}+\tilde{q}_{0}\left( \tilde{q}_{0}^{2}-M_{1}^{\ast
2}+M_{2}^{\ast 2}\right) I_{1}+\right.  \nonumber \\
&&\left. \frac{1}{4}\left( q^{2}\left( \left( M_{1}^{\ast }-M_{2}^{\ast
}\right) ^{2}-\tilde{q}_{0}^{2}\right) +\left( \tilde{q}_{0}^{2}-M_{1}^{\ast
2}+M_{2}^{\ast 2}\right) ^{2}\right) I_{0}\right] ,  \label{Pi33}
\end{eqnarray}%
\begin{eqnarray}
\mathop{\rm Im}\Pi _{R}^{11} &=&\mathop{\rm Im}\Pi _{R}^{22}=-\frac{1}{4\pi
q^{3}}\left[ tI_{2}+\tilde{q}_{0}\left( t-M_{1}^{\ast 2}+M_{2}^{\ast
2}\right) I_{1}\right.  \nonumber \\
&&\left. +\frac{1}{4}\left( 4q^{2}M_{1}^{\ast }M_{2}^{\ast }-q^{4}+\left( 
\tilde{q}_{0}^{2}-M_{1}^{\ast 2}+M_{2}^{\ast 2}\right) ^{2}\right) I_{0}%
\right] ,  \label{Pi11}
\end{eqnarray}%
\[
\mathop{\rm Im}\Pi _{R}^{A}=-\frac{1}{2\pi q}M_{1}^{\ast }M_{2}^{\ast
}I_{0}, 
\]%
\begin{equation}
\mathop{\rm Im}\Pi _{R}^{VA}=\frac{1}{8\pi q^{2}}\left[ 2tI_{1}+\left(
t-M_{1}^{\ast 2}+M_{2}^{\ast 2}\right) \tilde{q}_{0}I_{0}\right] .
\label{PiVA}
\end{equation}%
Here 
\begin{equation}
t\equiv \tilde{q}_{0}^{2}-q^{2},  \label{t}
\end{equation}%
and the functions $I_{n}$ are defined as 
\begin{eqnarray}
I_{n} &=&\tanh \left( \frac{q_{0}^{\prime }}{2T}\right) \int_{\varepsilon
_{0}}^{\infty }d\varepsilon _{2}\ \varepsilon _{2}^{n}\left[ \left(
1-f_{2}\left( \varepsilon _{2}\right) \right) f_{2}\left( \varepsilon
_{2}+q_{0}^{\prime }\right) \right.  \nonumber \\
&&\left. +f_{2}\left( \varepsilon _{2}\right) \left( 1-f_{2}\left(
\varepsilon _{2}+q_{0}^{\prime }\right) \right) \right] .  \label{Fn}
\end{eqnarray}%
with%
\begin{equation}
q_{0}^{\prime }=q_{0}-\mu _{1}+\mu _{2}.  \label{q0prime}
\end{equation}%
The lower cut-off in Eq.(\ref{Fn}) arises due to kinematical restrictions
and reads 
\begin{eqnarray}
\varepsilon _{0} &=&-\frac{\tilde{q}_{0}}{2t}\left( t-M_{1}^{\ast
2}+M_{2}^{\ast 2}\right)  \nonumber \\
&&-\frac{q}{2t}\sqrt{\left( t-\left( M_{2}^{\ast }+M_{1}^{\ast }\right)
^{2}\right) \left( t-\left( M_{2}^{\ast }-M_{1}^{\ast }\right) ^{2}\right) }.
\label{emin}
\end{eqnarray}%
This energy is real-valued if either $t>\left( M_{2}^{\ast }+M_{1}^{\ast
}\right) ^{2}$ or $t\leq \left( M_{1}^{\ast }-M_{2}^{\ast }\right)
^{2}\allowbreak $. The first case corresponds to creation of a real baryon
pair in the medium and requires inclusion of the vacuum (Feynman) piece of
polarizations for a correct calculation. As mentioned above, we do not
consider so large values of $t$. Thus, the imaginary part of the
polarization functions does not vanish for $t\leq \left( M_{1}^{\ast
}-M_{2}^{\ast }\right) ^{2}$. In terms of $q_{0}$ and $q$, this can be
written as 
\begin{equation}
\left( q_{0}-U_{1}+U_{2}\right) ^{2}\leq q^{2}+\left( M_{1}^{\ast
}-M_{2}^{\ast }\right) ^{2}.  \label{q0q}
\end{equation}%
Contrarily to the statement made in refs. (\cite{HPH95}, \cite{RPLP99}) that
the imaginary part of polarizations vanishes when the momentum transfer is
time-like, our calculation shows some domain of $q_{0}>q$ where the
imaginary part does not vanish even if we assume $M_{1}^{\ast }=M_{2}^{\ast
} $ (in fact, one can neglect the neutron-proton mass difference in nuclear
matter). This result is obviously correct, otherwise, the neutron decay
process $n\rightarrow p+e^{-}+\bar{\nu}$ , which requires $q_{0}^{2}-q^{2}>0$
for final leptons, would not be possible in neutron stars.

The nuclear matter in the core of neutron stars becomes transparent for
escaping neutrinos only at relatively low temperatures. Therefore we focus
on the low-temperature limit $\left( \mu _{B}/T\gg 1\right) $ and consider
degenerate baryons and leptons under thermal and $\beta $-equilibrium. In
this case the kinematical condition (\ref{emin}) restricts possible values
of the momentum transfer. Obviously, in the degenerate case the cut-off (\ref%
{emin}) of the integral (\ref{Fn}) should be lower than the baryon Fermi
energy%
\begin{eqnarray}
\varepsilon _{F2} &\geq &-\frac{\tilde{q}_{0}}{2t}\left( t-M_{1}^{\ast
2}+M_{2}^{\ast 2}\right)  \nonumber \\
&&-\frac{q}{2t}\sqrt{\left( t-\left( M_{2}^{\ast }+M_{1}^{\ast }\right)
^{2}\right) \left( t-\left( M_{2}^{\ast }-M_{1}^{\ast }\right) ^{2}\right) }.
\label{restr}
\end{eqnarray}

As mentioned above, the energy of the final lepton is close to its
individual Fermi energy $\mu _{l}$, while the condition of chemical
equilibrium is $\mu _{l}=\mu _{1}-\mu _{2}$, where the baryon chemical
potentials can be approximated by their individual Fermi energies. Thus, to
the lowest order in $T/\mu _{l}$ we can take $\tilde{q}_{0}=\varepsilon
_{F1}-\varepsilon _{F2}$, and $t=\left( \varepsilon _{F1}-\varepsilon
_{F2}\right) ^{2}-q^{2}$ in all smooth functions. Inserting this into Eq.(%
\ref{restr}) we obtain the following domain of the momentum transfer%
\begin{equation}
p_{F1}-p_{F2}<q<p_{F1}+p_{F2},  \label{qrestr}
\end{equation}%
where the imaginary part of the polarization functions does not vanish.

To the lowest accuracy in $T/\mu _{l}$, we obtain from (\ref{Fn}): \ 
\begin{equation}
I_{0}\simeq q_{0}^{\prime },\ \ \ \ \ I_{1}\simeq q_{0}^{\prime }\varepsilon
_{F2},\ \ \ \ \ I_{2}=q_{0}^{\prime }\varepsilon _{F2}^{2},  \label{Fd}
\end{equation}%
where $q_{0}^{\prime }\equiv \tilde{q}_{0}-\varepsilon _{F1}+\varepsilon
_{F2}$ is of the order of $T$.

By taking also $\tilde{q}_{0}=\varepsilon _{F1}-\varepsilon _{F2}$ in all
smooth functions of Eqs. (\ref{Pi00} - \ref{PiVA}), in the interval (\ref%
{qrestr}) we obtain 
\begin{equation}
\mathop{\rm Im}\Pi _{R}^{\mu \nu }=\frac{q_{0}^{\prime }}{8\pi q}\Phi ^{\mu
\nu },\ \ \ \ \mathop{\rm Im}\Pi _{R}^{VA}=\frac{q_{0}^{\prime }}{8\pi q}%
\Phi _{VA},\ \ \ \ \mathop{\rm Im}\Pi _{R}^{A}=\frac{q_{0}^{\prime }}{8\pi q}%
\Phi _{A},  \label{ImMFA}
\end{equation}%
with 
\begin{equation}
\Phi ^{00}=\left[ \left( \varepsilon _{F1}+\varepsilon _{F2}\right)
^{2}-q^{2}-\left( M_{1}^{\ast }-M_{2}^{\ast }\right) ^{2}\right] ,
\label{ImMF00}
\end{equation}%
\begin{eqnarray}
\Phi ^{03} &=&\Phi ^{30}=\frac{1}{q}\left[ \left( \varepsilon
_{F1}-\varepsilon _{F2}\right) \left( \left( \varepsilon _{F1}+\varepsilon
_{F2}\right) ^{2}-q^{2}\right) \right.  \nonumber \\
&&\left. -\left( \varepsilon _{F1}+\varepsilon _{F2}\right) \left(
M_{1}^{\ast 2}-M_{2}^{\ast 2}\right) \right] ,  \label{ImMF03}
\end{eqnarray}%
\begin{eqnarray}
\Phi ^{33} &=&\frac{1}{q^{2}}\left[ \left( \varepsilon _{F1}-\varepsilon
_{F2}\right) ^{2}\left( \left( \varepsilon _{F1}+\varepsilon _{F2}\right)
^{2}-q^{2}\right) \right.  \nonumber \\
&&-2\left( \varepsilon _{F1}^{2}-\varepsilon _{F2}^{2}\right) \left(
M_{1}^{\ast 2}-M_{2}^{\ast 2}\right)  \nonumber \\
&&\left. +q^{2}\left( M_{1}^{\ast }-M_{2}^{\ast }\right) ^{2}+\left(
M_{1}^{\ast 2}-M_{2}^{\ast 2}\right) ^{2}\right] ,  \label{ImMF33}
\end{eqnarray}%
\begin{equation}
\Phi ^{11}=\Phi ^{22}=\frac{1}{2q^{2}}\left[ q^{4}+4q^{2}\left( \varepsilon
_{F1}\varepsilon _{F2}-M_{1}^{\ast }M_{2}^{\ast }\right) -\left(
p_{F1}^{2}-p_{F2}^{2}\right) ^{2}\right] ,  \label{ImMF11}
\end{equation}%
\begin{equation}
\Phi _{A}=-4M_{1}^{\ast }M_{2}^{\ast },  \label{ImMFa}
\end{equation}%
\begin{equation}
\Phi _{VA}=\frac{1}{q}\left( \left( \varepsilon _{F1}-\varepsilon
_{F2}\right) \left( p_{F1}^{2}-p_{F2}^{2}\right) -\allowbreak \left(
\varepsilon _{F1}+\varepsilon _{F2}\right) q^{2}\right) .  \label{ImMFva}
\end{equation}

Contracting together the tensors, as indicated in Eq. (\ref{Qnu}), and
calculating the imaginary part to the lowest order in $T/\mu _{l}$ we obtain

\begin{eqnarray}
&&\mathop{\rm Im}\left( {\rm L}_{\alpha \beta }W^{\alpha \beta }\right) =%
\frac{q_{0}^{\prime }\omega _{1}}{\pi q}\left[ \left(
C_{V}^{2}+C_{A}^{2}\right) \left[ q_{0}\Phi ^{00}-2q\Phi ^{03}+q_{0}\Phi
^{33}\right. \right.  \nonumber \\
&&\left. +2q_{0}\Phi ^{11}+\left( q\Phi ^{00}-2q_{0}\Phi ^{03}+q\Phi
^{33}-2q\Phi ^{11}\right) \cos \theta \right]  \nonumber \\
&&\left. -2C_{A}^{2}\left( q_{0}-q\cos \theta \right) \Phi
_{A}-4C_{V}C_{A}\omega _{1}\left( q-q_{0}\cos \theta \right) \Phi _{VA} 
\right] \Psi (q_{0},q),  \label{ImTr}
\end{eqnarray}%
where $\theta $ is the angle between the ${\bf k}_{1}$ and ${\bf q}$
momenta. The step-function%
\[
\Psi (q_{0},q)\equiv \Theta \left( q+p_{F2}-p_{F1}\right) \Theta \left(
p_{F1}+p_{F2}-q\right) , 
\]%
$\allowbreak $ with $\Theta \left( x\right) =1$ if $x\geq 0$ and zero
otherwise, comes from the kinematical restrictions (\ref{qrestr}) and (\ref%
{q0q}). By inserting expression (\ref{ImTr}) into Eq. (\ref{Qnu}), and
taking $q_{0}\simeq \mu _{l}$, $q\simeq p_{Fl}$, we arrive to the integral
over $dq_{0}^{\prime }$ and the neutrino phase space. We omit this
straightforward calculation, which results in the expression given by Eq. (%
\ref{QMFA}). We stress here the fact that agreement between these two
methods is achieved only when we take into account non-conservation of the
baryon vector current.

In a non-relativistic limit $\left( P_{FB}\ll M_{B}^{\ast },\;\mu _{l}\ll
M_{B}^{\ast }\right) $, the leading terms of Eq. (\ref{QMFA}) give the
following expression 
\begin{equation}
Q_{L}=\frac{457\pi }{10080}G_{F}^{2}C^{2}\left( C_{V}^{2}+3C_{A}^{2}\right)
M_{1}^{\ast }M_{2}^{\ast }\mu _{l}T^{6}\Theta \left(
p_{Fl}+p_{F2}-p_{F1}\right) ,  \label{Qnr}
\end{equation}%
which coincides with the known non-relativistic result of Lattimer et al. %
\cite{Lat91}. As in that case, our formula (\ref{QMFA}) exhibits a threshold
dependence on the proton concentration. The ''triangle'' condition $%
p_{Fl}+p_{F2}\geq p_{F1}$, required by the step-function, is necessary for
conservation of the total momentum in the reaction. We pay attention,
however, that this condition should be supplemented by the one of chemical
equilibrium. As mentioned in introduction, in the relativistic regime, the
chemical equilibrium is possible only due to the fact that strong
interactions create a gap between the proton and neutron energy spectrums,
that is much larger than the mass difference of participating baryons.

The relative efficiency of the direct Urca processes involving different
kinds of baryons depends essentially on the composition of the $\beta $%
-stable nuclear matter in the core of a neutron star. In order to quantify
the relativistic effects in the direct Urca processes, we consider a
simplified model for degenerate nuclear matter of the standard composition
consisting on neutrons, protons and electrons under beta-equilibrium. We use
a Walecka-type \cite{Serot} self-consistent relativistic model of nuclear
matter, by assuming that the interactions among nucleons are mediated by
exchange of $\sigma ,\omega ,$ and $\rho $ mesons. In this model the baryon
effective mass $M^{\ast }=M-g_{\sigma B}\sigma _{0}$ as well as the
potential energies $U_{B}=g_{\omega B}\omega _{0}+g_{\rho B}t_{3B}b_{0}$ are
calculated in a self-consistent way. Here $\omega _{0}$, $b_{0}$, and $%
\sigma _{0}$ are, respectively, the mean-field values of the $\omega $, $%
\rho $, and $\sigma $-mesons; $g_{\omega B}$, $g_{\rho B}$, and $g_{\sigma
B} $ are the strong interaction coupling constants, and $t_{3B}$ is the
third component of isospin for the baryons. Parameters of the model are
chosen, as in ref. \cite{RPLP99}, to reproduce the nuclear matter
equilibrium density, the binding energy per nucleon, the symmetry energy,
the compression modulus, and the nucleon Dirac effective mass at saturation
density $n_{0}=0.16fm^{-3}$ . We have plotted in Fig. 1 the emissivity of
the Urca process, as given by our formula Eq. (\ref{QMFA}) in comparison
with the emissivity (\ref{Qnr}) given by Lattimer et al. Both magnitudes are
normalized with respect to the emissivity given by our formula at threshold
density. We can observe that relativistic effects dramatically modify the
emissivity. The non-relativistic approximation approaches our result only at
densities much smaller than the threshold density, indicated by an arrow on
the density axis. Due to the decrease of the nucleon effective mass, the
formula derived by Lattimer et al. predicts a decreasing of the emissivity
as density increases above the threshold. In contrast, our formula shows a
substantial increasing of the neutrino energy losses as we go to larger
densities. Of course, the exact value depends on the underlying nuclear
matter model.

Our Eq. (\ref{QMFA}) is obtained in the mean field approximation and does
not take into account correlations among the baryons, which normally have a
tendency to suppress neutrino energy losses. These calculations are out of
the scope of this letter, and we leave these unwieldy calculations for a
future work. At some densities neutrons may be superfluid and/or protons
superconducting. The direct Urca rate is then reduced. This can be taken
into account in the standard way (See e. g. \cite{ly94b}).

\begin{acknowledgments}
This work was supported by the Russian Foundation for Fundamental Research
Grant 00-02-16271 and by Spanish Grants DGES PB97- 1432 and AEN99-0692
\end{acknowledgments}

\vskip 0.3cm
\psfig{file=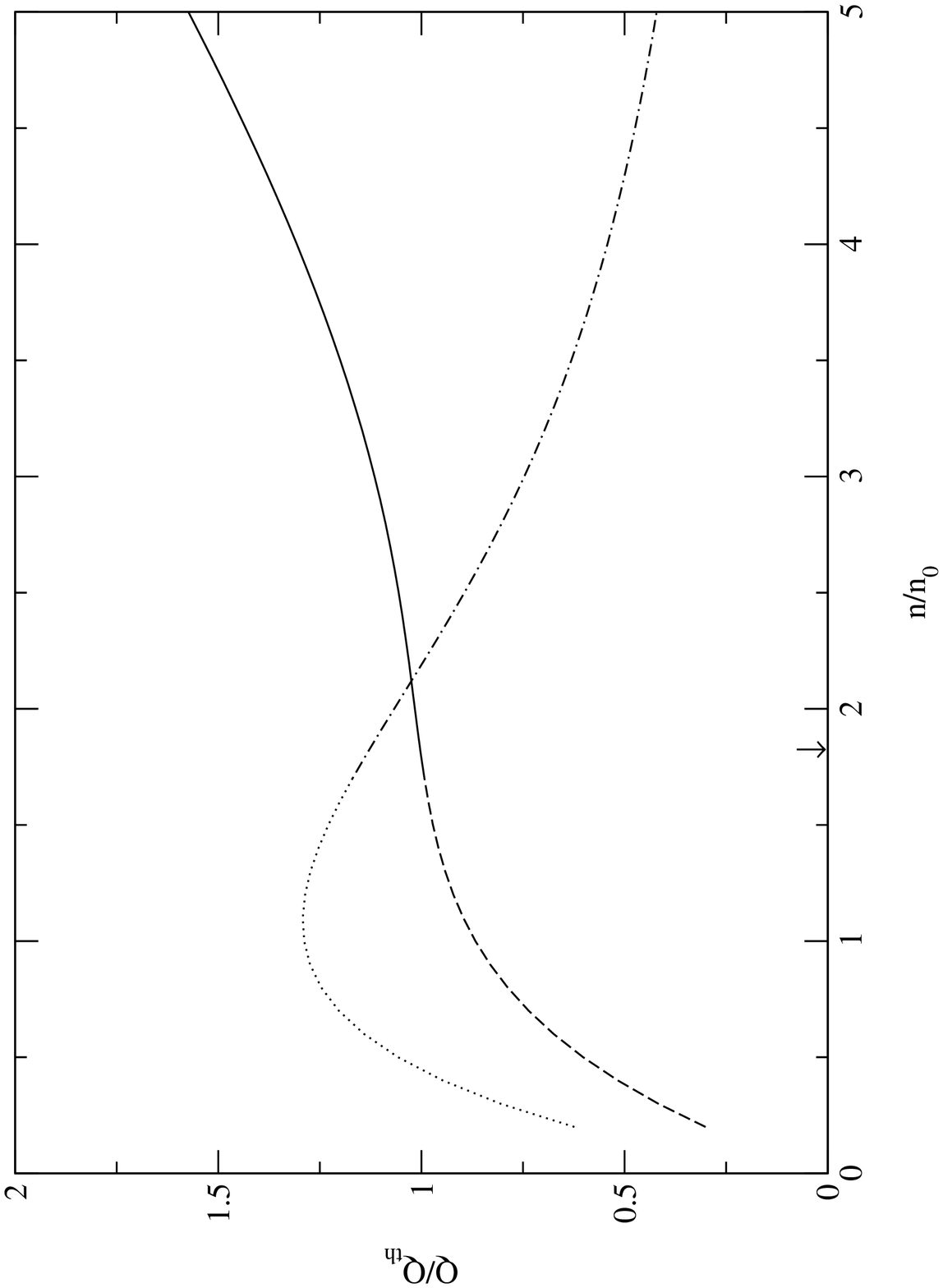}
Fig 1. Neutrino emissivity due to the Urca processes. We plot, as a dashed line
below the threshold density $n_{th}=1.79n_{0}$, and as a solid line above $%
n_{th}$ our result, given by Eq.(\ref{QMFA}). We also plot, for the sake of
comparison, the emissivity given by Lattimer et al, as a dotted line below $%
n_{th}$, and as a dashed-dotted line above $n_{th}$. Both magnitueds are
normalized with respect to the emissivity given by our formula at threshold
density. We labeled the threshold density by an arrow on the abscisa axis,
which shows the number density in units of saturation density $%
n_{0}=0.16fm^{-3}$.
\vskip 0.3cm
\end{document}